\documentclass[aps,preprint,prd,nofootinbib]{revtex4-1}
\pdfoutput=1	
\usepackage[utf8]{inputenc}
\usepackage{booktabs}
\usepackage{slashed}
\usepackage{indentfirst}
\usepackage[pdftex]{graphicx}
\usepackage{epstopdf}
\usepackage{amsfonts,amssymb,amsmath}
\usepackage{psfrag}

\usepackage{multirow}
\usepackage{hyperref}
\usepackage{mathrsfs} 
\usepackage{url}
\usepackage{appendix}
\usepackage{subfigure}

\newcommand{\GeV}{{\ \rm{GeV}}}

\graphicspath{{./pics/}}
\begin{document}

\title{Study of the dilepton electromagnetic decays of $\chi_{cJ}(1P)$}

\author{Xuan-He Wang\footnote[1]{These authors contributed equally to this work.}, Yue Jiang$^{[1]}$\footnote[1]{These authors contributed equally to this work.}, Tianhong Wang$^{[1]}$\footnote[2]{Corresponding author: Tianhong Wang, e-mail: thwang@hit.edu.cn.}, Xiao-Ze Tan$^{[1]}$, Geng Li$^{[1]}$, Guo-Li Wang$^{[1]}$ }

\address{Department of Physics, Harbin Institute of
	Technology, Harbin 150001, China}

\date{July 2, 2019}

\baselineskip=20pt

\vspace*{0.5cm}

\begin{abstract}

In this paper, the dilepton electromagnetic decays $\chi_{cJ}(1P) \to J/\psi e^+e^-$ and $\chi_{cJ}(1P) \to J\psi \mu^+\mu^-$, where $\chi_{cJ}$ denotes $\chi_{c0}$, $\chi_{c1}$ and $\chi_{c2}$, are calculated systematically in the improved Bethe-Salpeter method. The numerical results of decay widths and the invariant mass distributions of the final lepton pairs are given. The comparison is made with the recently measured experimental data of BESIII. It is shown that for the cases including $e^+e^-$, the gauge invariance is decisive and should be considered carefully. For the processes of $\chi_{cJ}(1P) \to J/\psi e^+e^-$, the branching fraction are: $\mathcal{B}[\chi_{c0}(1P) \to J/\psi e^+e^-]=1.06^{+0.16}_{-0.18} \times 10^{-4}$, $\mathcal{B}[\chi_{c1}(1P) \to J/\psi e^+e^-]=2.88^{+0.50}_{-0.53} \times 10^{-3}$, and $\mathcal{B}[\chi_{c2}(1P) \to J/\psi e^+e^-]=1.74^{+0.22}_{-0.21} \times 10^{-3}$. The calculated branching fractions of $\chi_{cJ}(1P)\to J/\psi \mu^+\mu^-$ channels are: $\mathcal{B}[\chi_{c0}(1P) \to J/\psi \mu^+\mu^-]=3.80^{+0.59}_{-0.64} \times 10^{-6}$, $\mathcal{B}[\chi_{c1}(1P) \to J/\psi \mu^+\mu^-]=2.04^{+0.36}_{-0.38} \times 10^{-4}$, and $\mathcal{B}[\chi_{c2}(1P) \to J/\psi \mu^+\mu^-]=1.66^{+0.19}_{-0.19} \times 10^{-4}$.

 \vspace*{0.5cm}


\end{abstract}

\maketitle


\section{INTRODUCTION}

The dilepton electromagnetic (EM) decays, $M \to M_f \ell^+\ell^-$, where $M$ and $M_f$ are initial and final mesons, respectively, are of significance in revealing the structure of hadronic states and the mechanisms of the interactions between hadrons and electromagnetic field \cite{Landsberg:1986fd,Fu:2011yy,Li:2009wz}. As the EM decays are much cleaner than hadronic decays, dilepton EM decay processions have been well studied in the light quark sector for years \cite{Tanabashi:2018oca}. In Ref. \cite{Faessler:1999de}, the dilepton decays of unflavored light mesons $\rho$, $\omega$, $\phi$, $\eta$, $\eta^{\prime}$, $\pi^0$, $f_0$ and $a_0$ are calculated to give the dilepton spectra, which can be used to provide references for experimental searches of such decays. However, there is still little study of such decays in the charm and bottom sectors. Recently, both BESIII \cite{Ablikim:2017kia,Ablikim:2019jqp} and LHCb \cite{Aaij:2017vck} have observed the dilepton EM decays of $\chi_{cJ}$ and $\psi(3686)$, where $\chi_{cJ}$ refers to $\chi_{c0}$, $\chi_{c1}$ and $\chi_{c2}$. By analysing the cascade decays of $\psi(3686) \to \chi_{cJ} e^+e^-$, $\chi_{cJ} \to J/\psi\gamma$ and $\psi(3686) \to \chi_{cJ}\gamma$, $\chi_{cJ} \to J/\psi e^+e^-$, the BESIII Collaboration measured the branching fractions of $\chi_{cJ} \to J/\psi e^+e^-$, which are $(1.51 \pm 0.30 \pm 0.13) \times 10^{-4}$, $(3.73 \pm 0.09 \pm 0.25) \times 10^{-3}$, $(2.48 \pm 0.08 \pm 0.16) \pm 10^{-3}$ \cite{Ablikim:2017kia}. This has been gradually bringing more attention on such decays.

The partial width of dilepton EM decays can be obtained with the transition form factors of charmonia, which can be calculated from QCD models. Similar to the form factors, the partial width is $Q$-dependent, where $Q$ is the invariant mass of the final leptons. Based on this, the spectra of invariant mass could be derived, which can provide more information of the process and the inner structure of charmonia involved. Meanwhile, the calculation of dilepton EM decays are deeply related to radiative decays. When we take the limit $Q=0$, the transition form factors of this two electromagnetic decays should share the same value, which are constrained by the Ward identity. In our previous researches, the radiative decay channels have been well calculated for multiple charm mesons, like $\chi_{c1}$, $X(3872)$ \cite{Wang:2010ej}, $B_c$ \cite{Ju:2015mns}, etc. In Ref. \cite{Wang:2010ej}, the radiative decay channels were calculated for $\chi_{c1}(2P)$ as a charmonium candidate of $X(3872)$, which has been confirmed to be a P-wave charmonium by PDG.

The P-wave triplet states $\chi_{cJ}$ are mainly produced by the radiative decay of $\psi(3686)$ \cite{Tanabashi:2018oca}. Unlike $1^{--}$ charmonium $J/\psi$ and $\psi(3686)$, $\chi_{cJ}$ are rarely produced directly in the $e^+e^-$ collisions. For charmonia below the $D\bar{D}$ threshold, like $\psi(3686)$ and $\chi_{cJ}(1P)$, the electromagnetic decay modes become important. However, as the charmonia discovered next to $J/\psi$ and $\psi(3686)$, the decays of $\chi_{cJ}(1P)$ are relatively less learned. Thus, theoretical calculations of the EM decays of $\chi_{cJ}(1P)$ based on QCD model may provide more information of the inner structure and mechanism of charmonia. In this paper, we use the improved Bethe-Salpeter (BS) method \cite{Salpeter:1951sz,Salpeter:1952ib} to calculate the transition amplitude. Recently, the relativistic correction within this formalism is also presented \cite{Kim:2003ny,Wang:2013lpa}. This method is widely applied to the calculation of heavy meson physics \cite{Tan:2018lao,Geng:2018qrl}. The triplet states $\chi_{cJ}(1P)$ are under the $D\bar{D}$ threshold and have no OZI-allowed hadronic decays. Thus, EM decays, especially the radiative process has rather large contribution. In Ref. \cite{Luchinsky:2017pby}, the ratios between the branching fractions of $\chi_{cJ}\to J/\psi \ell^+\ell^-$ processes and the corresponding radiative decay channels are calculated by assuming the virtual photons are on-shell, making the transition form factors constant in the calculation. Here we will not use this assumption, but calculate the decay widths and branching fractions of the dilepton EM decay processes with the full range form factors and hope to provide more comparable results.

This paper is organized as follows. In Sec. \ref{formalism}, we present the theoretical formalism based on improved BS method and give the form of the invariant amplitude and partial width. In Sec. \ref{numerical_results}, we present the numerical results and make comparison with the experimental data. Finally, we give the conclusions are provided Sec. \ref{summary}. Some details of the Salpeter wave functions are presented in the Appendix.

\section{THE FORMALISM}\label{formalism}

\begin{figure}[htbp]
	\centering
	\subfigure[]{\label{xc1_fey}
		\includegraphics[width=0.45\textwidth]{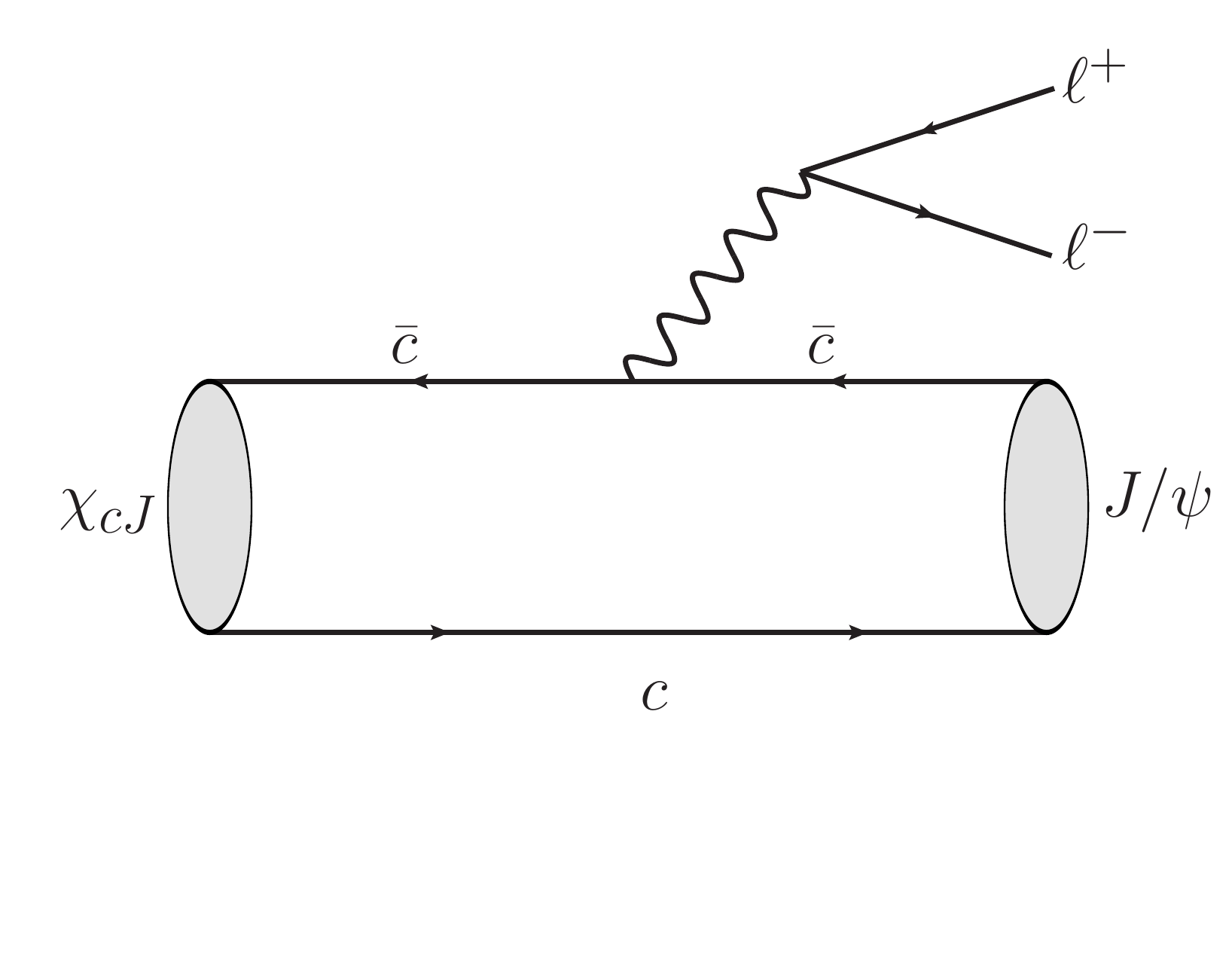}}
	\qquad
	\subfigure[]{\label{xc1_fey_anti}
		\includegraphics[width=0.45\textwidth]{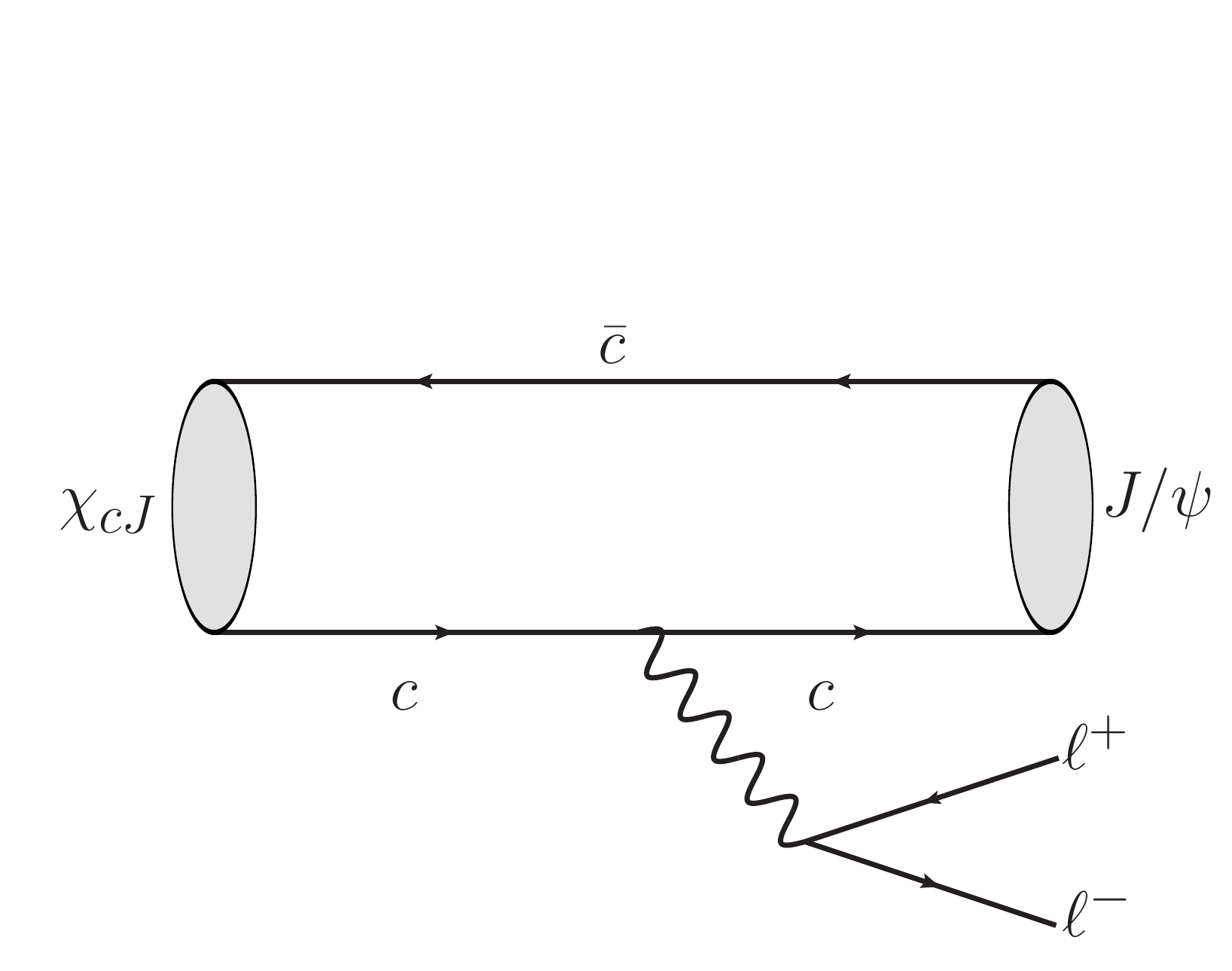}}
	\caption{The dilepton Feynman diagrams for the $\chi_{cJ} \rightarrow J/\psi\ell^+\ell^-$ transition. }\label{dil_feynman}
	\vspace{-1em}
\end{figure}

Considering the Feymann diagram Fig. \ref{dil_feynman}, the invariant amplitude of $\chi_{cJ} \to J/\psi \ell^+\ell^-$ can be written as:
\begin{align}\label{amp_dilep}
\mathcal{M} = \dfrac{\mathrm{i} e_q}{Q^2} \bar{u}_{\ell^-} \gamma_\mu v_{\ell^+} \langle J/\psi | \gamma^\mu | \chi_{cJ} \rangle ,
\end{align}
where we have defined the invariant mass $Q=\sqrt{(P_2+P_3)^2}$, with $P_2$ and $P_3$ being the momenta of the final negative-charged and positive-charged lepton, respectively; $e_q = \dfrac{2}{3} e \ (e_{\bar{q}} = - \dfrac{2}{3} e)$ is the charge of the charm quark (anti-quark). Within Mandelstam formalism, the hadronic matrix element can be expressed as an overlapping integral over the Salpeter wave functions of the initial and final mesons \cite{Wang:2010ej}:
\begin{eqnarray}\label{h_matrix_E}
\begin{aligned}
\langle J/\psi | \gamma^\mu | \chi_{cJ} \rangle &= \int \dfrac{\mathrm{d}^3 q_\bot}{(2 \pi)^3} Tr \Big\{ \dfrac{\slashed{P}}{M} \bar{\varphi}_{P_f}^{'++} (q^{'}_{2 \bot} ) \gamma^\mu \varphi_{P}^{++} (q_\bot) \\
&\quad - \overline{\varphi}_{P_f}^{'++} (q^{'}_{1 \bot}) \dfrac{\slashed{P}}{M} \varphi_{P}^{++} (q_\bot) \gamma^\mu \Big\} ,
\end{aligned}
\end{eqnarray}
where $\varphi^{++}$ and $\varphi^{'++}$ are the positive energy parts of Salpeter wave functions of the initial and final heavy mesons, respectively; $q$, $q^{\prime}_1$ and $q_2^{\prime}$ are the relative momenta between quark and anti-quark in the initial and final mesons, respectively; $P$ and $P_f$ are the momenta of the initial and final mesons, respectively. $M$ is the mass of the initial meson. We also use the definitions:
\begin{eqnarray}
\begin{aligned}
&q_{\bot}^{\mu}=q^{\mu}-\frac{(P\cdot q)}{M^2}P^{\mu} ,\\
& q^\prime_{1 \bot} = q_\bot - \alpha_1^\prime P_{f \bot} \quad (\alpha_1^\prime \equiv \dfrac{m_1^\prime}{m_1^\prime + m_2^\prime} = \dfrac{1}{2}) ,  \\
& q^\prime_{2 \bot} = q_\bot + \alpha_2^\prime P_{f \bot} \quad (\alpha_2^\prime \equiv \dfrac{m_2^\prime}{m_1^\prime + m_2^\prime} = \dfrac{1}{2}) .
\end{aligned} 
\end{eqnarray}
where $m_1^\prime$ and $m_2^\prime$ are the masses of quark and anti-quark of the final meson, respectively.

The basic method to derive the explicit wave functions is to take the instantaneous approximation to simplify the original BS equation. This could reduce a 4-dimensional BS equation to 3-dimensional solvable Salpeter equations. Furthermore, as the negative energy part of the wave function in the Salpeter equations has a rather small contribution \cite{Wang:2011jt}, only the positive energy part is included in our calculation. The explicit form of wave functions involved in our calculation are directly given in the appendix. For interested readers, more detailed processes of deriving the functions as well as solving the instantaneous BS equation can be found in our previous papers \cite{Wang:2007av,Wang:2009er,Wang:2005qx}. 

The hadronic matrix element is reduced to several form factors after finishing the overlap integrals. Here we give the final forms of the hadronic matrix elements representing the cases of $0^{++} \to 1^{--}$, $1^{++} \to 1^{--}$ and $2^{++} \to 1^{--}$ respectively.
\begin{eqnarray}\label{form_matrix_E1}
\begin{aligned}
\left\langle J/\psi \mid \gamma^\mu \mid \chi_{c0} \right\rangle &= P^{\mu}(P\cdot\varepsilon_{f}) t_{1}+P_{f}^{\mu}(P\cdot\varepsilon_{f})t_{2}+\varepsilon_{f}^{\mu}t_{3}, \\
\left\langle J/\psi \mid \gamma^\mu \mid \chi_{c1} \right\rangle &= \epsilon^{\mu\xi\alpha\beta}\varepsilon_{1\xi}P_{\alpha}P_{f\beta}(P\cdot\varepsilon_{f})s_{1} \\
&+ P^{\mu}\epsilon^{\xi\nu\alpha\beta}\varepsilon_{1\xi}\varepsilon_{f\nu}P_{\alpha}P_{f\beta}s_2  \\
&+ P_f^{\mu}\epsilon^{\xi\nu\alpha\beta}\varepsilon_{1\xi}\varepsilon_{f\nu}P_{\alpha}P_{f\beta}s_3+\epsilon^{\mu\xi\nu\alpha}\varepsilon_{1\xi}\varepsilon_{f\nu}P_{\alpha}s_4, \\
\left\langle J/\psi \mid \gamma^\mu \mid \chi_{c2} \right\rangle &=P^{\mu}(P\cdot\varepsilon_{f})P_f^{\xi}P_f^{\nu}\varepsilon_{\xi\nu} u_1+P^{\mu}P_f^{\xi}\varepsilon_{f}^{\nu}\varepsilon_{\xi\nu} u_2 \\
&+P_f^{\mu}(P\cdot\varepsilon_f) P_f^{\xi}P_f^\nu\varepsilon_{\xi\nu} u_3+P_f^{\mu}P_f^\xi\varepsilon_{f}^\nu\varepsilon_{\xi\nu} u_4  \\
&+\varepsilon_f^{\mu}P_f^\xi P_f^\nu \varepsilon_{\xi\nu} u_5+P_{f\xi}\varepsilon^{\mu\xi}(P\cdot\varepsilon_f) u_6+\varepsilon_{f\xi}\varepsilon^{\mu\xi} u_7.
\end{aligned}
\end{eqnarray}
Here, $t_i$, $s_i $ and $u_i$ are form factors. The polarization tensor of $\chi_{c2}$ is represented by $\varepsilon_{\xi\nu}$. $\varepsilon_1$ and $\varepsilon_f$ denote the polarization vectors of $\chi_{c1}$ and $J/\psi$, respectively. Note that in Eq. (\ref{form_matrix_E1}), the form factors are not independent. Due to the Ward identity $(P^{\mu}-P_f^{\mu})\left\langle J/\psi \mid \gamma_\mu \mid \chi_{cJ} \right\rangle\Big|_{Q=0}=0$, they are related by the following constraint conditions:
\begin{eqnarray}\label{ward_id}
\begin{aligned}
t_3&=-\dfrac{1}{2} (M^2-M_f^2)(t_1+t_2),\\
s_4&=-\dfrac{1}{2} (M^2-M_f^2)(s_1+s_2),\\
u_7&=\dfrac{1}{2} (M^2-M_f^2)(u_2+u_4).
\end{aligned}
\end{eqnarray} 
After using the constraint conditions to replace $t_3, s_4$ and $u_7$, the hadronic matrix are parameterized with the left form factors.

Summed up over the polarization, the squared amplitude becomes
\begin{align}\label{form_matrix}
\sum|\mathcal{M}|^2 = \dfrac{4}{9}\dfrac{(4\pi)^2\alpha^2}{Q^2}l_{\mu\nu}h^{\mu\nu} . 
\end{align}
The leptonic and hadronic tensor $l_{\mu\nu}$ and $h^{\mu\nu}$ takes the following forms:
\begin{eqnarray}
\begin{aligned}
l_{\mu\nu} &=-\dfrac{Q^2}{2}g^{\mu\nu}+4(P_2^\mu P_3^\nu+P_2^\nu P_3^\mu), \\
h^{\mu\nu} &= \sum \left\langle J/\psi \mid \gamma^\mu \mid \chi_{cJ} \right\rangle\left\langle \chi_{cJ} \mid \gamma^{\nu\dagger} \mid J/\psi \right\rangle , \\
&=y_1P^\mu P^\nu +y_2(P^\mu P_f^\nu+P^\nu P_f^\nu)+y_3P_f^\mu P_f^\nu +y_4g^{\mu\nu},
\end{aligned}
\end{eqnarray}
where $y_i$ are functions that depend on the invariant mass $Q$. The explicit form of $y_i$ are derived by summing up the polarization vector(tensor) using following formulae:
\begin{eqnarray}
\begin{aligned}
\sum_{\lambda}\varepsilon^{\mu}_{(\lambda)}\varepsilon^{\nu}_{(\lambda)} &= g_{\bot}^{\mu\nu}, \\
\sum_{\lambda}\varepsilon^{\mu\nu}_{(\lambda)}\varepsilon^{\rho\sigma}_{(\lambda)} &= \dfrac{1}{2}(g^{\mu\sigma}_{\bot}g^{\nu\rho}_{\bot}+g^{\mu\rho}_{\bot}g^{\nu\sigma}_{\bot}) - \dfrac{1}{3}g^{\mu\nu}_{\bot}g^{\rho\sigma}_{\bot},
\end{aligned}
\end{eqnarray}
where $g^{\mu\nu}_{\bot} = \dfrac{P^{\mu}P^{\nu}}{M^2}-g^{\mu\nu}$ is defined. Here we give $y_i$ for the $\chi_{c0}\to J/\psi \ell^+\ell^-$ case as an example:
\begin{eqnarray}
\begin{aligned}
y_1 &=\dfrac{(M^2+M_f^2-Q^2)^2t_1^2}{4M_f^2}-M_f^2t_1t_2,\\
y_2 &=-\dfrac{(M^2-M_f^2)(M^2+M_f^2-Q^2)t_1^2}{4M_f^2}+\dfrac{(M^-M_f^2)t^2_2}{2}-\dfrac{Q^2(M^2+3M_f^2+Q^2)t_1t_2}{4M_f^2},\\
y_3 &= \dfrac{(M^-M_f^2)^2t^2_1}{4M_f^2}+\bigg[ \dfrac{(Q^2-2M_f^2)^2}{4M_f^2}-M^2 \bigg] t_2^2-\dfrac{(M^2-M_f^2)(2M_f^2-Q^2)t_1t_2}{2M_f^2},\\
y_4 &= -\dfrac{(M^2-M_f^2)^2}{4}(t1+t2)^2.
\end{aligned}
\end{eqnarray}

The decay width of a three-body process is given by 
\begin{eqnarray}\label{decay_width}
\begin{aligned}
\Gamma = \frac{1}{2M}\int\frac{\mathrm{d}^3P_f}{(2\pi)^32E_f}\frac{\mathrm{d}^3P_2}{(2\pi)^32E_2}&\frac{\mathrm{d}^3P_3}{(2\pi)^32E_3}(2\pi)^4\delta^{(4)}(P-P_f-P_2-P_3)\overline{|\mathcal{M}|^2} .\\
\end{aligned} 
\end{eqnarray}
The invariant mass spectra of final leptons has the form:
\begin{align}\label{3_body_decay}
\dfrac{\mathrm{d}\Gamma}{\mathrm{d}Q} &= \frac{1}{256\pi^3M^3}\frac{1}{Q}\sqrt{\lambda(M^2,M_f^2,Q^2)\lambda(Q^2,M_2^2,M_3^2)}\int\mathrm{d}\cos\theta\overline{|\mathcal{M}|^2} , 
\end{align}
where $\theta$ is the angle between the final lepton and the meson in the center of mass of the lepton pair. $\lambda$ has the form: 
\begin{align}
\lambda(x,y,z) &= x^2+y^2+z^2-2xy-2yz-2xz . 
\end{align}

\section{NUMERICAL RESULTS AND DISCUSSIONS}\label{numerical_results}

In previous papers, we have solved the corresponding full Salpeter equations for different mesons. In Ref. \cite{Chang:2010kj}, we fixed the parameters in the model by fitting the mass spectra of charmonia and bottomonia. In this paper, we use the same parameter values as those in Ref. \cite{Chang:2010kj}. The masses of mesons involved are listed as follows:
\begin{eqnarray}
\begin{aligned}
&M_{J/\psi} = 3.097 \GeV, &\quad M_{\chi_{c0}} = 3.415 \GeV,   \\
&M_{\chi_{c1}} = 3.511 \GeV, &M_{\chi_{c2}} = 3.556 \GeV. 
\end{aligned}
\end{eqnarray}

\begin{figure}[htbp]
\centering
\subfigure[$\chi_{c0} \rightarrow J/\psi e^+ e^-$.]{\label{fig_chi_c0_ee}
	\includegraphics[width=0.415\textwidth]{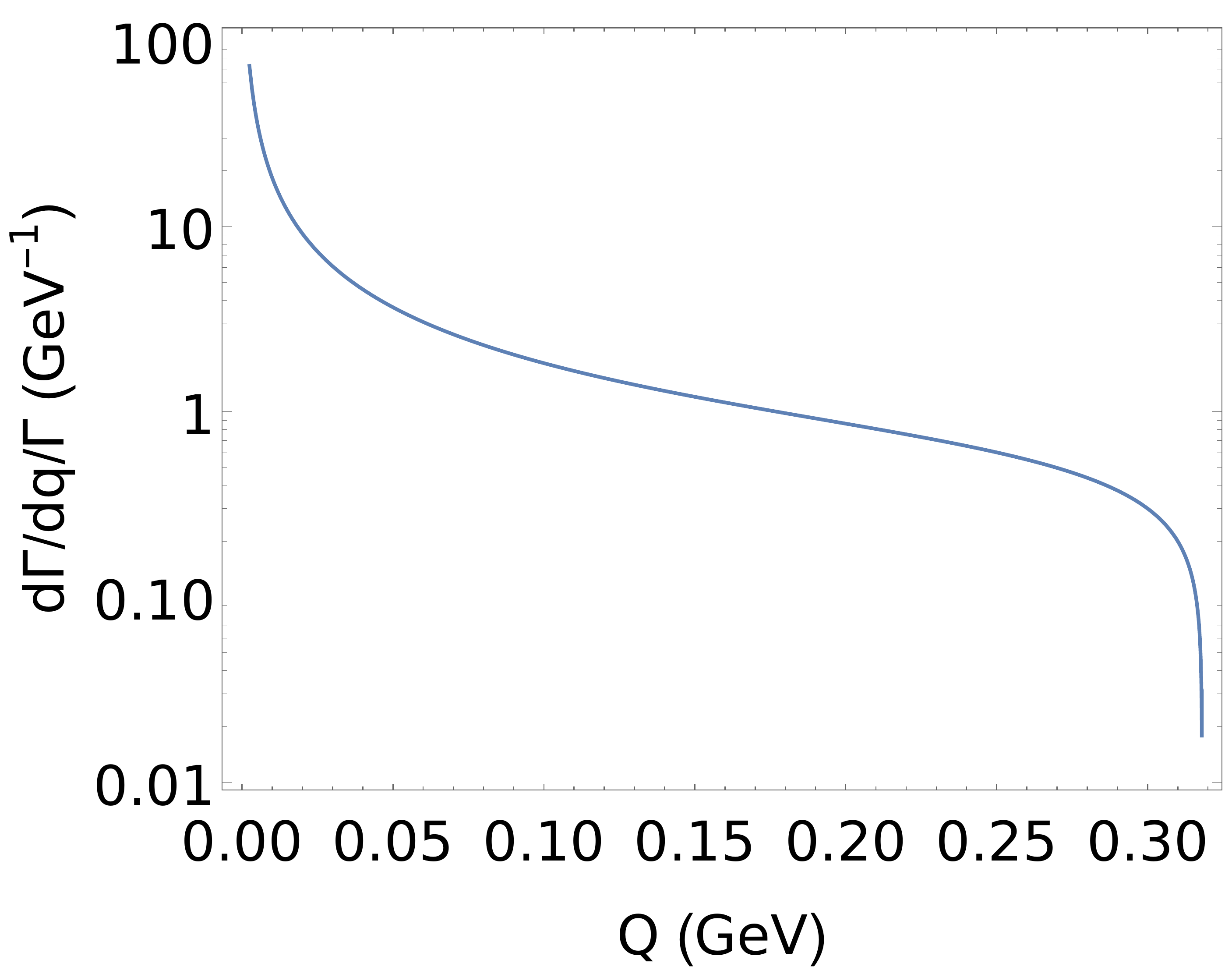}}
\qquad
\subfigure[$\chi_{c1} \rightarrow J/\psi e^+ e^-$.]{\label{fig_chi_c1_ee}
	\includegraphics[width=0.415\textwidth]{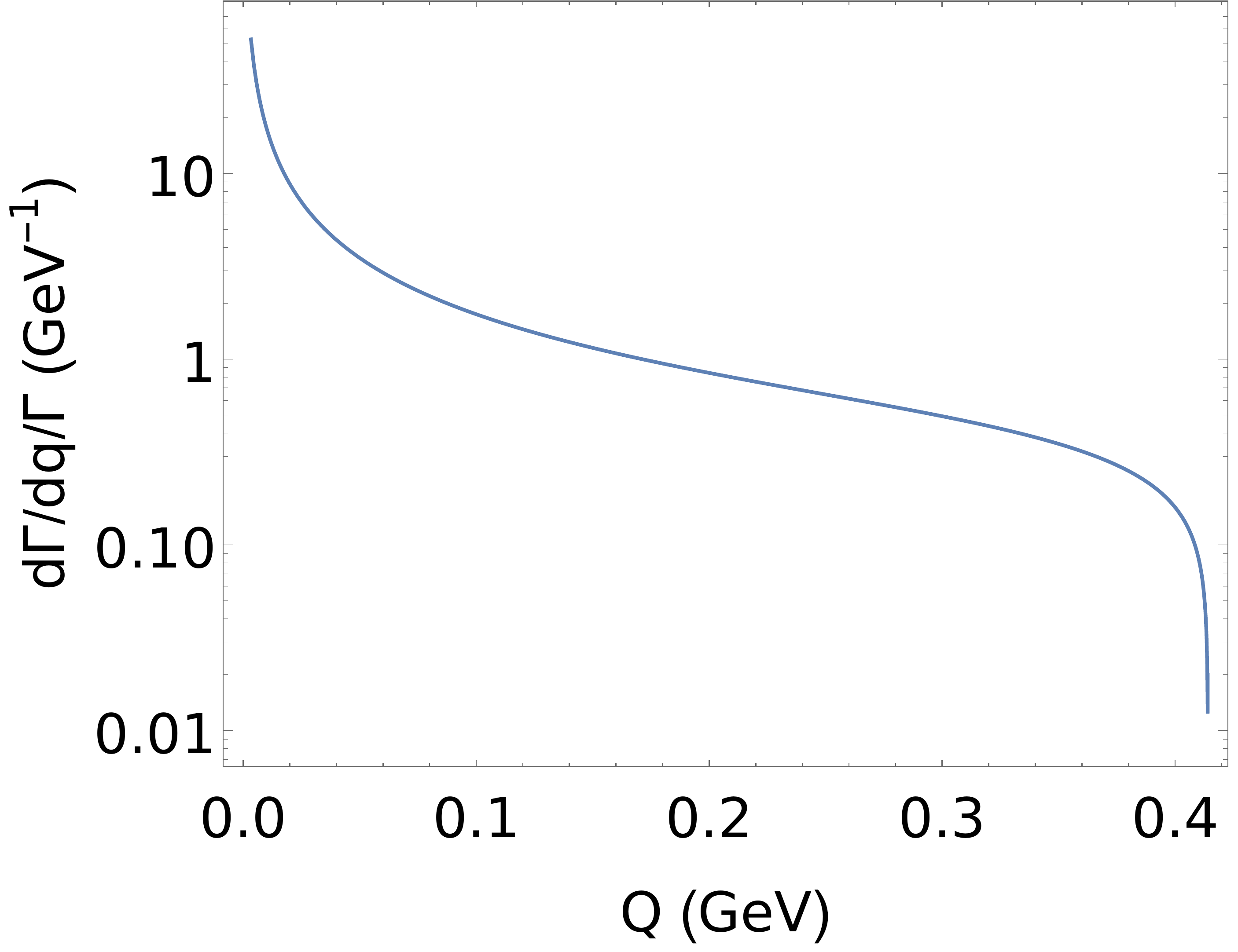}}
\subfigure[$\chi_{c2} \rightarrow J/\psi e^+ e^-$.]{\label{fig_chi_c2_ee}
	\includegraphics[width=0.415\textwidth]{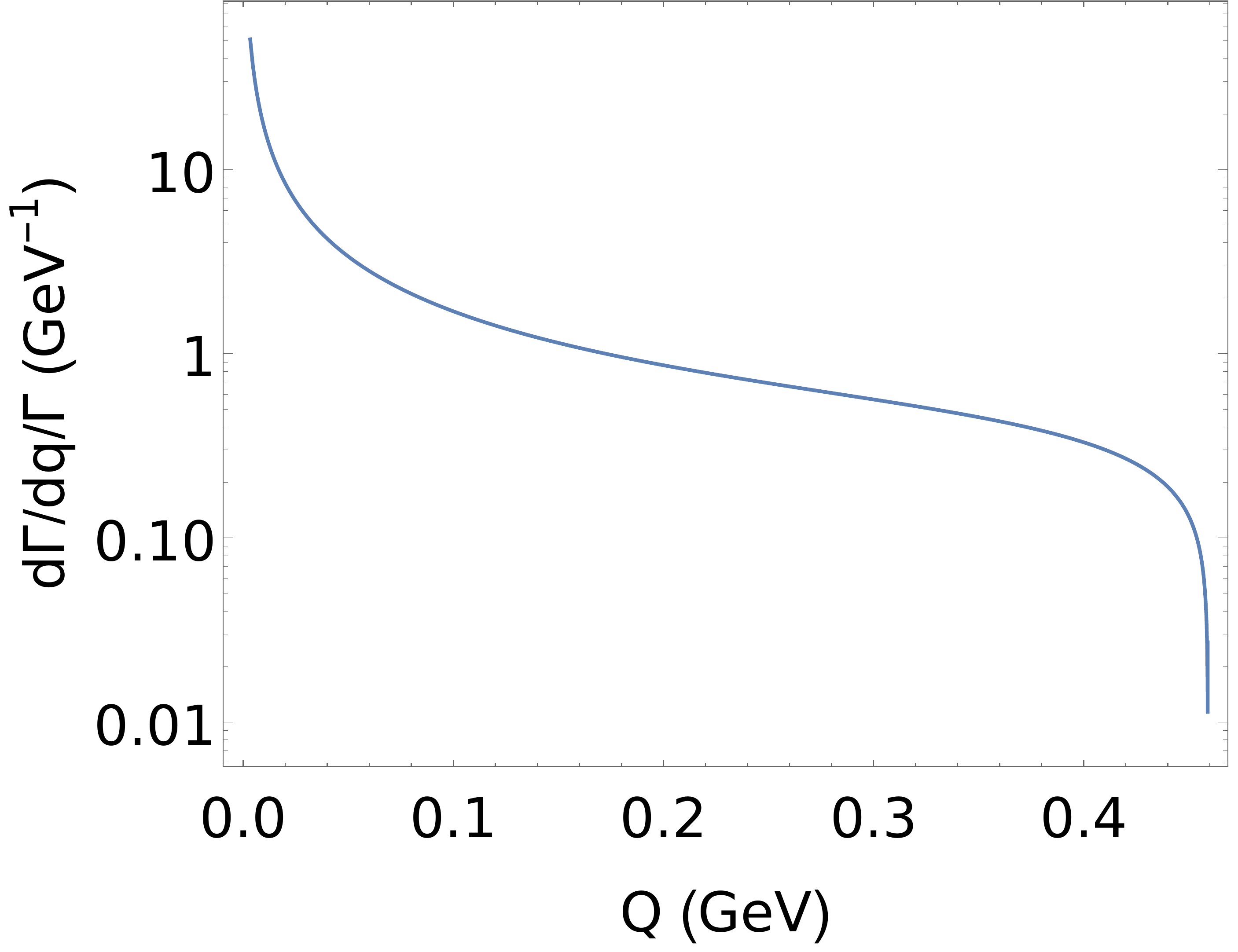}}
\qquad
\subfigure[The spectrum for $\chi_{c1} \rightarrow J/\psi e^+ e^-$ by BESIII \cite{Ablikim:2017kia}. ]{\label{chi_c1_e_exp}
	\includegraphics[width=0.42\textwidth]{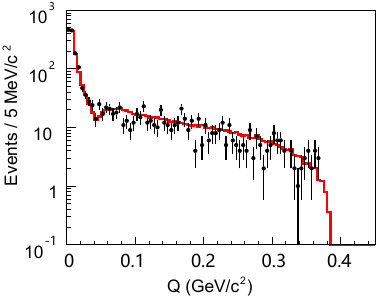}}
\subfigure[The spectrum for $\chi_{c2} \rightarrow J/\psi e^+ e^-$ by BESIII \cite{Ablikim:2017kia}. ]{\label{chi_c2_e_exp}
	\includegraphics[width=0.42\textwidth]{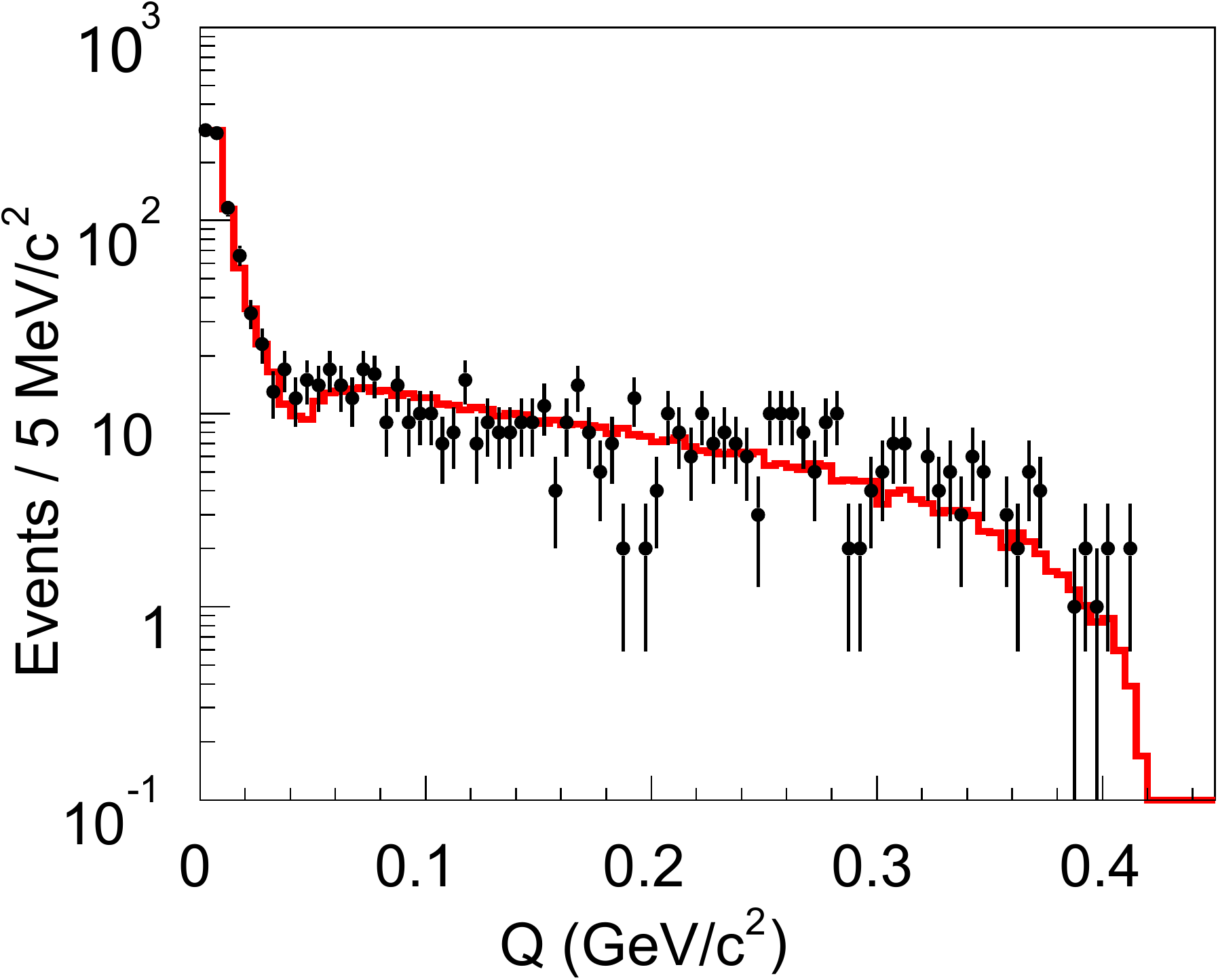}}
\caption{The invariant mass spectra for the decays $\chi_{cJ} \to J/\psi e^+ e^-$. The experimental data for the decays $\chi_{c1,2} \to J/\psi e^+ e^-$ is shown in (d) and (e), where the histograms are for the signal Monte-Carlo (MC) simulation. }\label{spc_ee}
\vspace{-1em}
\end{figure}

By fitting the mass spectra of charmonia, the mass of constituent charm quarks are set to $m_i = 1.62\GeV$ in our calculation. The mass spectra of lepton pair of $\chi_{cJ} \to J/\psi e^+e^-$ processes are shown in Fig. \ref{spc_ee}. The horizontal axis represents the invariant mass $Q \equiv \sqrt{P_2^2+P_3^2}$, and the vertical axis represents the partial width. For each calculated spectrum, a peak occurs near the lower limit of invariant mass $Q\equiv \sqrt{P_2^2+P_3^2}=2m_e$, similar to the spectra given by BESIII, whose vertical axis represents the number of events instead. This could be understood from Eq. (\ref{form_matrix}), where the squared amplitude is proportion to $1/Q^4$, causing it to increase rapidly at the lower limit. Besides, since the mass of electron is four orders of magnitudes smaller than those of the charmonia, the distribution of decay width gets a large contribution when the invariant mass is small enough. This indicates that the value of the form factors near the $Q=2m_e$ can bring large contribution to the results in our calculation. Thus, it is necessary to increase the accuracy of the calculated form factors at this range. At the range where the invariant mass become larger, the curve flattens as the effect from phase space becomes dominant. 

In Table \ref{tb_electron}, our numerical results of decay widths and branching fractions are given. The uncertainties are achieved by varying the (anti-)quark masses and parameters in the interaction potential by $\pm10\%$.According to our numerical results, the $\chi_{c2} \to J\psi e^+e^-$ process has the largest decay width while the $\chi_{c0} \to J\psi e^+e^-$ process has the smallest. Meanwhile, the calculated branching fractions of $\chi_{c0,1} \to J/\psi e^+e^-$ processes are consistent with the experimental data in the given uncertainties, while the calculated result of $\chi_{c2} \to J/\psi e^+e^-$ is comparable. 

\begin{table}[htbp]
\centering
\caption{The decay widths and branching fractions of dilepton EM decays of $\chi_{cJ}$ with $e^+e^-$ as final leptons. }\label{tb_electron}
\begin{tabular}{lccc}
\toprule[1.5pt]
\qquad Decay mode \qquad & \quad $\Gamma (\rm{keV})$ \quad & \quad $\mathcal{B}$  ($\times 10^{-3}$) \quad & \quad $\mathcal{B}_{\mathrm{EXP}}$  ($\times 10^{-3}$) \quad \\
\midrule[1pt]
$\chi_{c0} \rightarrow J/\psi e^+ e^-$ & \quad $1.14^{+0.17}_{-0.19}$ & $0.106^{+0.016}_{-0.018}$ & 0.151$\pm 0.030 \pm 0.013$ \cite{Ablikim:2017kia} \\
$\chi_{c1} \rightarrow J/\psi e^+ e^-$ & \quad $2.42^{+0.42}_{-0.44}$ & $2.88^{+0.50}_{-0.53}$ & 3.73$\pm 0.09 \pm 0.25$ \cite{Ablikim:2017kia} \\
$\chi_{c2} \rightarrow J/\psi e^+ e^-$ & \quad $3.42^{+0.43}_{-0.40}$ & $1.74^{+0.22}_{-0.21}$ & 2.48$\pm 0.08 \pm 0.16$ \cite{Ablikim:2017kia} \\
\bottomrule[1.5pt]
\end{tabular}
\end{table}

\begin{table}[htbp]
	\centering
	\caption{The decay widths and branching fractions of dilepton EM decays of $\chi_{cJ}$ with $\mu^+\mu^-$ as final leptons. }\label{tb_muon}
	\begin{tabular}{lccc}
		\toprule[1.5pt]
		\qquad Decay mode \qquad & \quad $\Gamma  (\rm{keV})$ \quad & \quad $\mathcal{B}$  ($\times 10^{-4}$) \quad & \quad $\mathcal{B}_{\mathrm{EXP}}$  ($\times 10^{-4}$) \quad \\
		\midrule[1pt]
		$\chi_{c0} \rightarrow J/\psi \mu^+ \mu^-$ & \quad $(4.10^{+0.64}_{-0.69})\times 10^{-2}$ & $(3.80^{+0.59}_{-0.64}) \times 10^{-2}$ & $< 0.2$ \cite{Ablikim:2019jqp} \\
		$\chi_{c1} \rightarrow J/\psi \mu^+ \mu^-$ & \quad $0.171^{+0.031}_{-0.032}$ & $2.04^{+0.36}_{-0.38}$ & 2.51$\pm 0.18 \pm 0.20$ \cite{Ablikim:2019jqp} \\
		$\chi_{c2} \rightarrow J/\psi \mu^+ \mu^-$ & \quad $0.327^{+0.037}_{-0.037}$ & $1.66^{+0.19}_{-0.19}$ & 2.33$\pm 0.18 \pm 0.29$ \cite{Ablikim:2019jqp} \\
		\bottomrule[1.5pt]
	\end{tabular}
\end{table}

\begin{table}[htbp]
	\centering
	\caption{The ratios of branching fractions of $\chi_{cJ} \to J/\psi \mu^+\mu^-$ and $\chi_{cJ} \to J/\psi e^+e^-$. }\label{tb_ratio}
	\begin{tabular}{lccc}
		\toprule[1.5pt]
		\qquad Decay mode\qquad & \quad This work \quad & \quad BESIII \cite{Ablikim:2019jqp} & 	Luchisky \cite{Luchinsky:2017pby} \quad \\
		\midrule[1pt]
		$\dfrac{\mathcal{B}[\chi_{c0} \to J/\psi \mu^+\mu^-]}{\mathcal{B}[\chi_{c0} \to J/\psi e^+e^-]}(\times 10^{-2})$&\quad  $3.60^{+0.01}_{-0.01}$ & $<14$ & $2.72$\\
		\\[-4mm]
		$\dfrac{\mathcal{B}[\chi_{c1} \to J/\psi \mu^+\mu^-]}{\mathcal{B}[\chi_{c1} \to J/\psi e^+e^-]}(\times 10^{-2})$  &\quad $7.08_{-0.02}^{+0.03}$ &\quad $6.73\pm0.51\pm0.50$ & $5.93$\\
		\\[-4mm]
		$\dfrac{\mathcal{B}[\chi_{c2} \to J/\psi \mu^+\mu^-]}{\mathcal{B}[\chi_{c2} \to J/\psi e^+e^-]}(\times 10^{-2})$ &\quad $9.54_{-0.08}^{+0.06}$ &\quad $9.40\pm0.79\pm1.15$ & $7.36$ \\
		\bottomrule[1.5pt]
	\end{tabular}
\end{table}
 
 \begin{figure}[htbp]
 	\centering
 	\subfigure[$\chi_{c0} \rightarrow J/\psi \mu^+ \mu^-$.]{\label{fig_chi_c0_mu}
 		\includegraphics[width=0.40\textwidth]{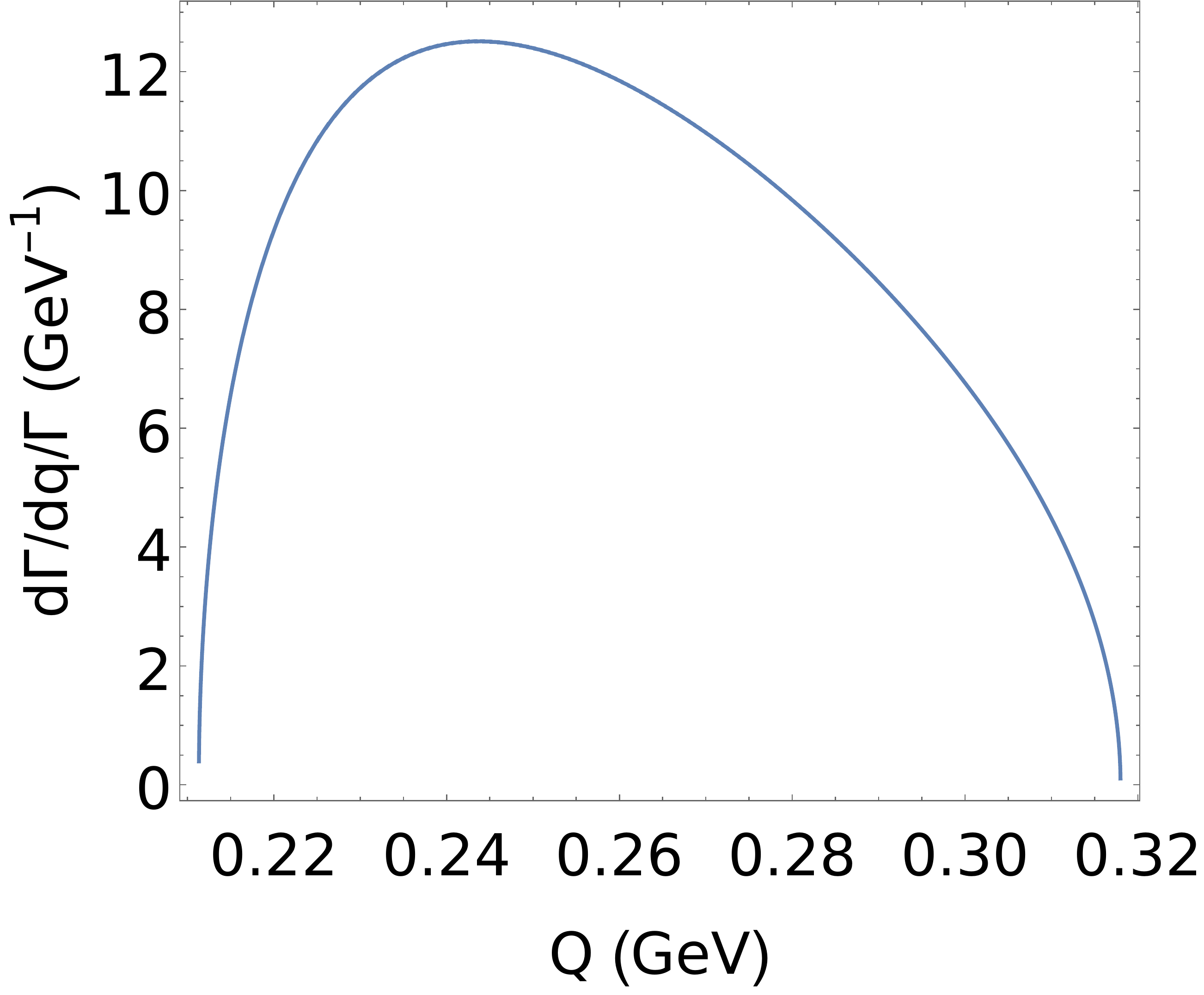}}
 	\qquad
 	\subfigure[$\chi_{c1} \rightarrow J/\psi \mu^+ \mu^-$.]{\label{fig_chi_c1_mu}
 		\includegraphics[width=0.37\textwidth]{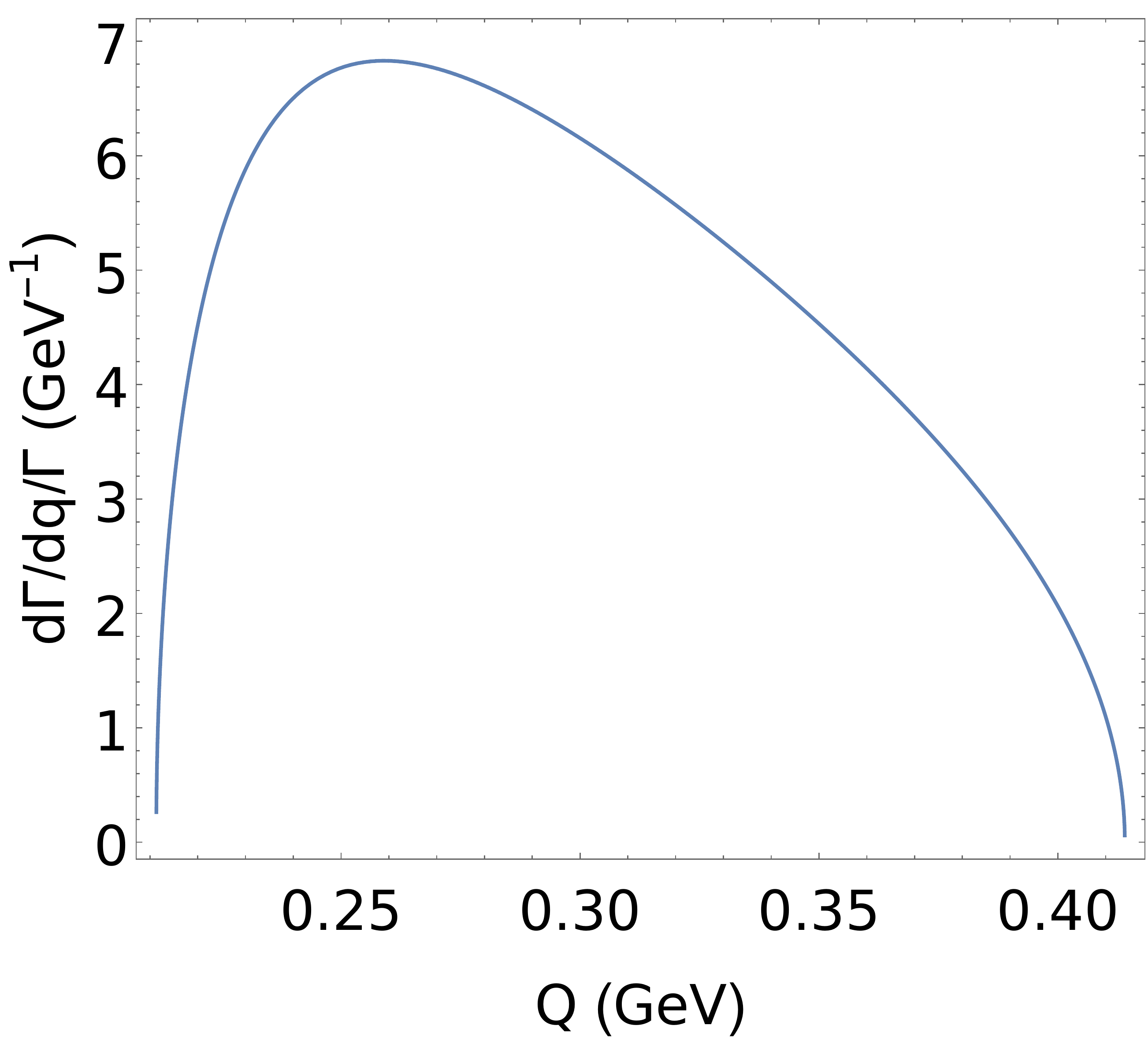}}
 	\subfigure[$\chi_{c2} \rightarrow J/\psi \mu^+ \mu^-$.]{\label{fig_chi_c2_mu}
 		\includegraphics[width=0.37\textwidth]{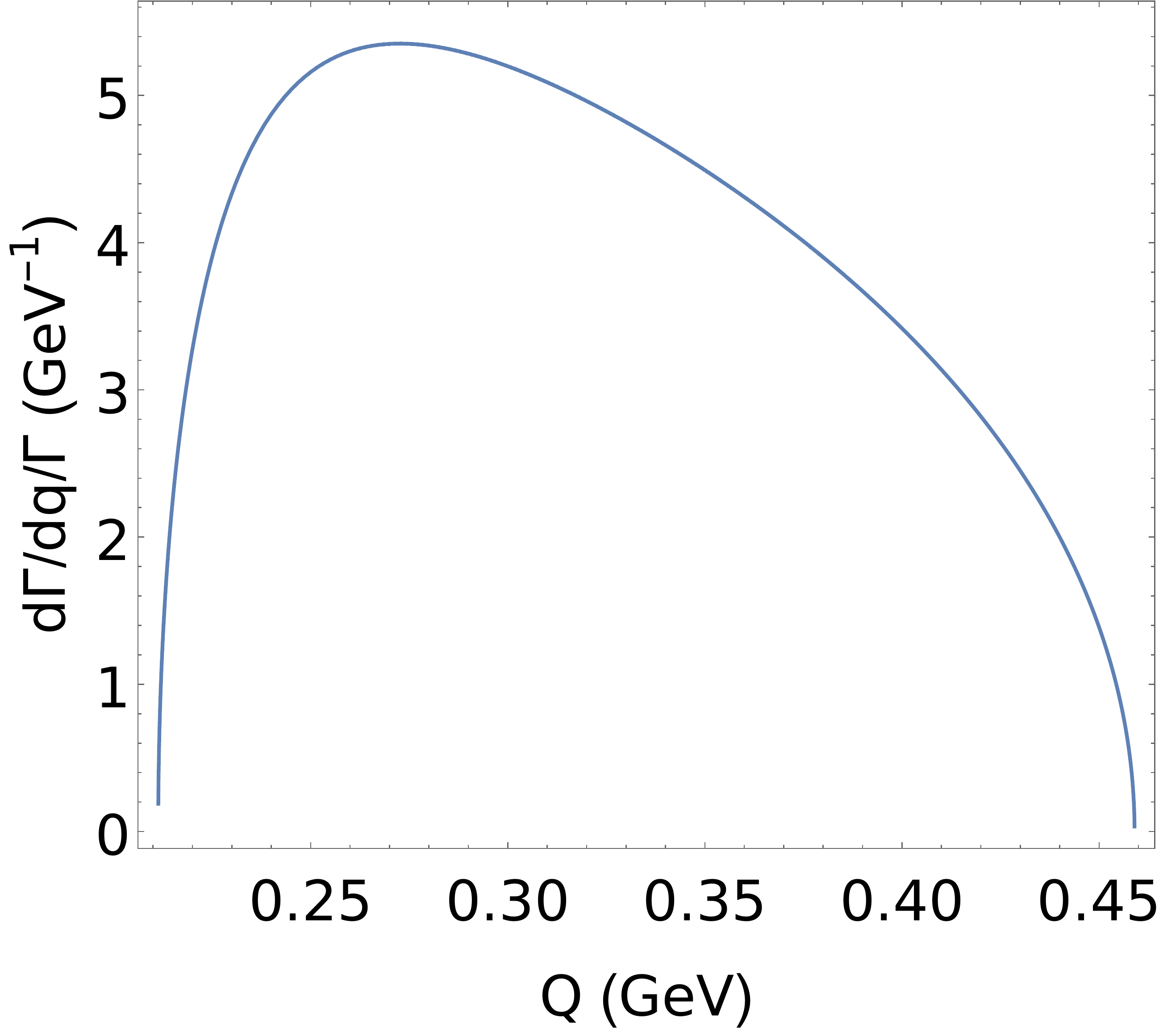}}
 	\qquad
 	\subfigure[The spectrum for $\chi_{c1} \rightarrow J/\psi \mu^+ \mu^-$ by BESIII \cite{Ablikim:2019jqp}. ]{\label{chi_c1_mu_exp}
 		\includegraphics[width=0.38\textwidth]{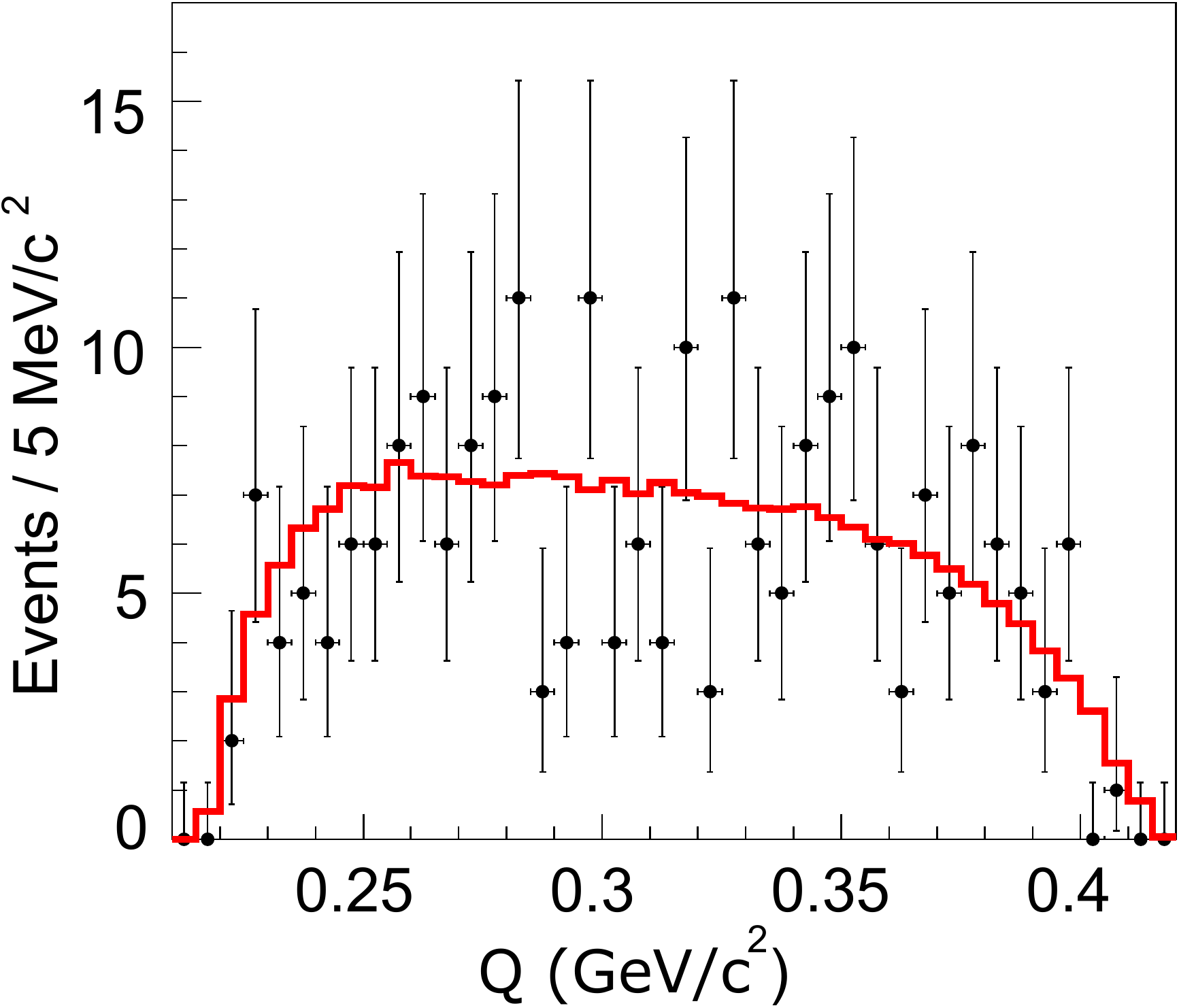}}
 	\subfigure[The spectrum for $\chi_{c2} \rightarrow J/\psi \mu^+ \mu^-$ by BESIII \cite{Ablikim:2019jqp}.]{\label{chi_c2_mu_exp}
 		\includegraphics[width=0.38\textwidth]{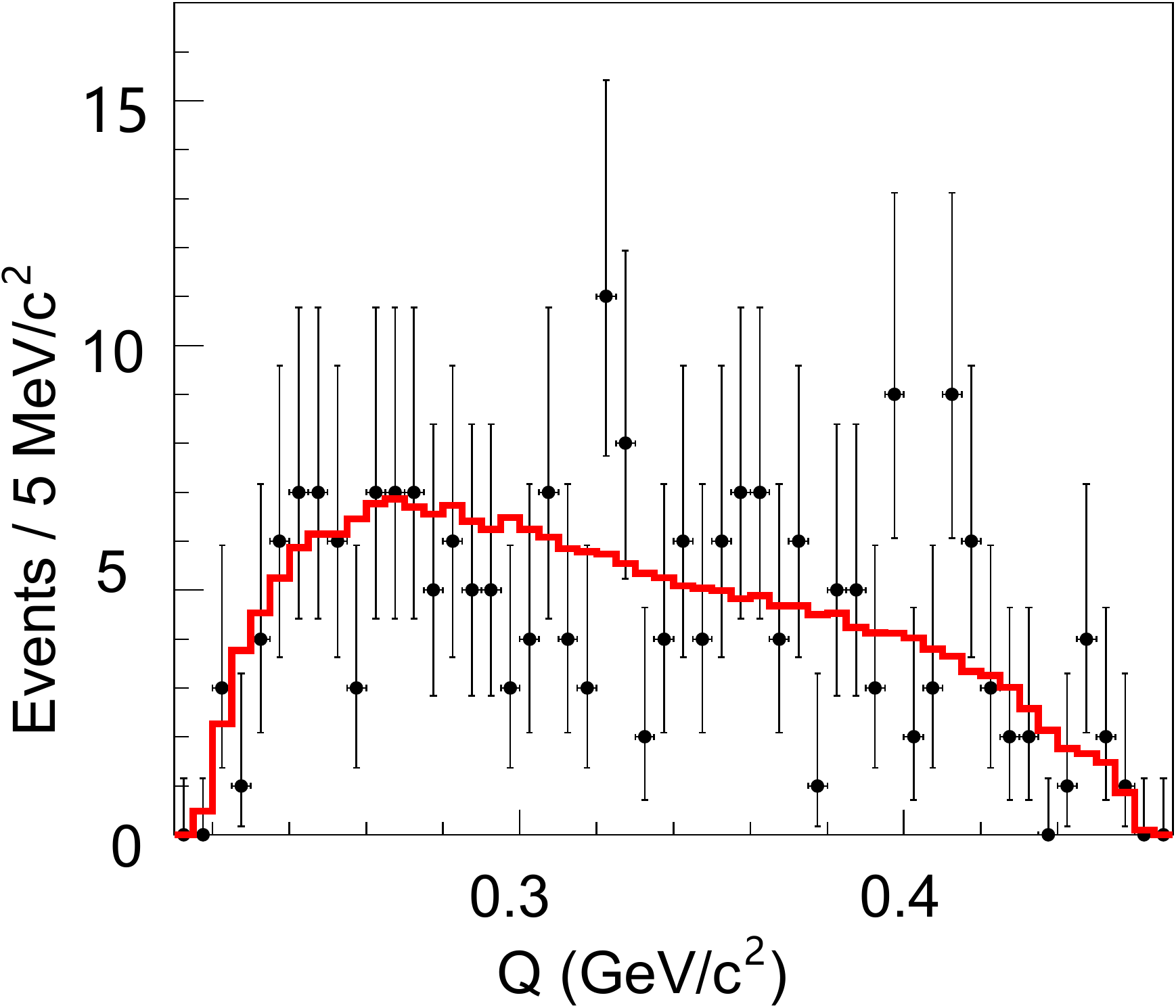}}
 	\qquad
 	\subfigure[The spectrum for $\chi_{c1,2} \rightarrow J/\psi \mu^+ \mu^-$ by LHCb \cite{Aaij:2017vck}.]{\label{muon_lhcb}
 	 		\includegraphics[width=0.42\textwidth]{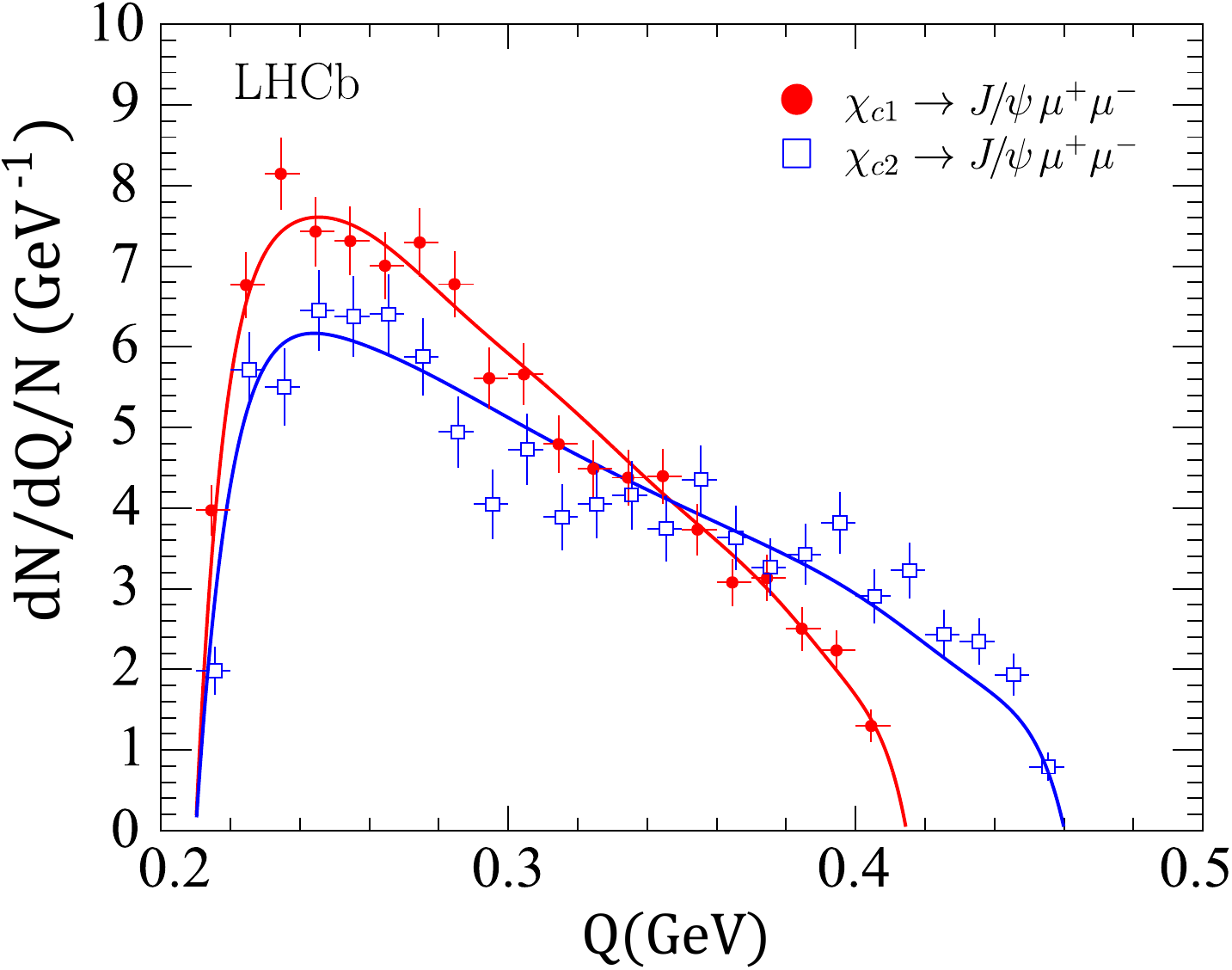}}
 	\caption{The invariant mass spectra for the decays $\chi_{cJ} \to J/\psi \mu^+ \mu^-$. The experimental data for the decays $\chi_{c1,2} \to J/\psi \mu^+ \mu^-$ is shown in (d), (e) and (f), where the histograms are for the signal MC simulation. }\label{spc_mu}
 	\vspace{-1em}
 \end{figure}
 
In Fig. \ref{spc_mu}, the invariant mass spectra of the muon pairs are shown along with the results of BESIII \cite{Ablikim:2019jqp} and LHCb \cite{Aaij:2017vck}. The axes of spectra given by BESIII represent number of events while the axis of the spectrum given by LHCb represents partial width. The curves given by LHCb are results of simulation, which uses the model described in Ref. \cite{Faessler:1999de}. Our results are consistent with the experimental data qualitatively. Clearly, as the mass of a muon is much larger than that of an electron, the peaks appeared at about $0.25\GeV$ in the muon spectra are not so sharp. Compared to the experimental data, our muon spectra shares the same lower and upper limit of invariant mass, while our peak are slightly lower than the spectra given by LHCb. Our results of decay widths and branching fractions for the $\chi_{cJ} \to J/\psi \mu^+\mu^-$ process are shown in Table \ref{tb_muon}. The experimental branching fractions given by BESIII \cite{Ablikim:2019jqp} are also listed to make comparison. For the process of $\chi_{c0} \to J/\psi \mu^+\mu^-$, the experiment has only given the upper limit of branching fraction. Our calculated result stays within the given limits. The results of $\chi_{c1} \to J/\psi \mu^+\mu^-$ also consist with those of the experiments. Though the central value of $\chi_{c2} \to J/\psi \mu^+\mu^-$ process is slightly smaller, it is still consistent with the experimental data after the uncertainties being considered. 

Furthermore, the ratios between branching fractions are given in Table \ref{tb_ratio}. The theoretical uncertainties have been reduced. The central values of our results of $\chi_{c1,2}$ are slightly larger than those of the experimental data provided by BESIII, while our result of $\chi_{c0}$ is under the upper limit of experimental data. Generally, the branching fractions of $\chi_{cJ} \to J/\psi \mu^+\mu^-$ are about one order of magnitude smaller than that of $\chi_{cJ} \to J/\psi e^+e^-$. In Ref. \cite{Luchinsky:2017pby}, the same processions are calculated with a relation between dilepton decays and radiative decays. The relation was given by Ref. \cite{Faessler:1999de},  in which the transition form factors was derived based on the Vector Meson Dominance (VMD) model within the constraints from quark counting rules. Generally, their results are relatively smaller to the central values of experimental data given by BESIII.
 
\section{SUMMARY}\label{summary}

In this work, the dilepton EM decays between $\chi_{cJ}$ and $J/\psi$ is studied with the improved Bethe-Salpeter method. Our results were compared with the recent BESIII experiments, the distribution of $\chi_{c1,2} \to J/\psi \ell^+\ell^-$ are consistent with the experimental spectra qualitatively, showing the suitability of our method. For the process of $\chi_{c0,1} \to J/\psi e^+e^-$ and $\chi_{cJ} \to J/\psi \mu^+\mu^-$, our calculated branching fractions consist with experiments under the given uncertainties, while the result of $\chi_{c2} \to J/\psi e^+e^-$ channel is comparable. Furthermore, the ratios of branching fractions are given to reduce theoretical uncertainties. The ratios have given good agreements with the experimental data of BESIII. So far, the explicit data of the involved channels has only been measured by BESIII, further comparison could be made with more experiment in the future.

\section*{ACKNOWLEDGEMENTS}

This work was supported in part by the National Natural Science Foundation of China (NSFC) under Grants No. 11405037, No. 11575048 and No. 11505039.


\section*{APPENDIX: Bethe-Salpeter Wave Function}
In the previous section, we have discussed how to derive the invariant amplitude and decay width. But to get the numerical results of a specific decay channel, the explicit Salpeter wave functions of the mesons involved are necessary. 

For the $0^+$ ($^3P_0$) state, the positive energy part of the wave function takes the form \cite{Wang:2007av}:
\begin{equation}
\varphi^{++}_{0+}(q_\bot) = A_1(q_\bot) + \dfrac{\slashed{P}}{M} A_2(q_\bot) + \dfrac{\slashed{q}_\bot}{M} A_3(q_\bot) + \dfrac{\slashed{P} \slashed{q}_\bot}{M^2} A_4(q_\bot), 
\end{equation}
where
\begin{align}
&A_1 = \dfrac{(\omega_1+\omega_2)q_\bot^2}{2(m_1\omega_2+m_2\omega_1)}\Big[f_{a1}+\dfrac{m_1+m_2}{\omega_1+\omega_2}f_{a2}\Big], \notag\\
&A_2 = \dfrac{(m_1-m_2)q_\bot^2}{2(m_1\omega_2+m_2\omega_1)}\Big[f_{a1}+\dfrac{m_1+m_2}{\omega_1+\omega_2}f_{a2}\Big], \notag\\
&A_3 = \dfrac{M}{2}\Big[f_{a1}+\dfrac{m_1+m_2}{\omega_1+\omega_2}f_{a2}\Big], \notag\\
&A_4 = \dfrac{M}{2}\Big[\dfrac{\omega_1+\omega_2}{m_1+m_2}f_{a1}+f_{a2}\Big].
\end{align}
In our expressions above, $P$ and $M$ denote the momentum and mass of the meson, while $q$ denotes the relative momenta of the quarks. $m_1$ ($m_2$) and $\omega_1$ ($\omega_2$) denotes the mass and energy of the constituent quark (anti-quark), respectively. Note that we have $m_1=m_2$ for the charmonia. These representation are used likewise in our expressions below. $f_{ai}$ ($i=1,2$) are the radial wave functions, which are obtained by solving the Salpeter equations of the $0^+$ state.

For the $1^{++}$ ($^3P_1$) state, the positive energy part of the wave function is \cite{Wang:2007av}:
\begin{equation}
\varphi^{++}_{1^{++}}(q_\bot) = \mathrm{i} \epsilon_{\mu \nu \alpha \beta} \dfrac{P^\nu}{M} q^\alpha_\bot \varepsilon_1^\beta \gamma^\mu \bigg[  B_1(q_\bot) + \dfrac{\slashed{P}}{M} B_2(q_\bot) + \dfrac{\slashed{q}_\bot}{M} B_3(q_\bot) + \dfrac{\slashed{P} \slashed{q}_\bot}{M^2} B_4(q_\bot)   \bigg] ,
\end{equation}
where
\begin{align}
& B_1 = \dfrac{1}{2} \Big[ f_{b1} + \dfrac{\omega_1 + \omega_2}{m_1 + m_2} f_{b2} \Big] , \notag \\
& B_2 = - \dfrac{1}{2} \Big[ \dfrac{m_1 + m_2}{\omega_1 + \omega_2} f_{b1} + f_{b2} \Big] , \notag \\
& B_3 = \dfrac{M(\omega_1-\omega_2)}{m_1 \omega_2 + m_2 \omega_1} B_1 , \notag \\
& B_4 = - \dfrac{M(m_1 + m_2)}{m_1 \omega_2 + m_2 \omega_1} B_1 .
\end{align}
$f_{bi}$ ($i=1,2$) are the radial wave functions obtained by solving the Salpeter equations of the $1^{++}$ state. 

The positive energy part of the wave function of the $2^+$ $(^2P_3)$ state is written as \cite{Wang:2009er}:
\begin{align}
\varphi^{++}_{2^{+}} &= \quad\varepsilon_{\mu\nu}q^{\mu}_{\bot}q^{\nu}_{\bot}\bigg[C_1(q_{\bot}) + \dfrac{\slashed{P}}{M}C_2(q_{\bot})+\dfrac{\slashed{q}_{\bot}}{M}C_3(q_{\bot}) + \dfrac{\slashed{P}\slashed{q}_{\bot}}{M^2}C_4(q_{\bot})\bigg], \notag\\
&+ M\varepsilon_{\mu\nu}\gamma^{\mu}q_{\bot}^{\nu}\bigg[C_5(q_{\bot}) + \dfrac{\slashed{P}}{M}C_6(q_{\bot})+\dfrac{\slashed{q}_{\bot}}{M}C_7(q_{\bot}) + \dfrac{\slashed{P}\slashed{q}_{\bot}}{M^2}C_8(q_{\bot})\bigg] , 
\end{align}
where
\begin{align}
&C_1 = \dfrac{1}{2M(m_1\omega_2+m_2\omega_1)}\big[(\omega_1+\omega_2)q_{\bot}^2f_{c3}+(m_1+m_2)q_{\bot}^2f_{c4} + 2M^2\omega_2f_{c5} - 2M^2m_2f_{c6}\big], \notag \\
&C_2 = \dfrac{1}{2M(m_1\omega_2+m_2\omega_1)}\big[(m_1-m_2)q_{\bot}^2f_{c3}+(\omega_1-\omega_2)q_{\bot}^2f_{c4} + 2M^2m_2f_{c5} - 2M^2\omega_2f_{c6}\big], \notag \\
&C_3 = \dfrac{1}{2}\bigg[f_{c3}+\dfrac{m_1+m_2}{\omega_1+\omega_2}f_{c4}-\dfrac{2M^2}{m_1\omega_2+m_2\omega_1}f_{c6}\bigg], \notag \\
&C_4 = \dfrac{1}{2}\bigg[\dfrac{\omega_1+\omega_2}{m_1+m_2}f_{c3}+f_{c4}-\dfrac{2M^2}{m_1\omega_2+m_2\omega_1}f_{c5}\bigg], \notag \\
&C_5 = \dfrac{1}{2}\bigg[f_{c5}-\dfrac{\omega_1+\omega_2}{m_1+m_2}f_{c6}\bigg], \qquad\qquad C_6=\dfrac{1}{2}\bigg[-\dfrac{m_1+m_2}{\omega_1+\omega_2}f_{c5}+f_{c6}\bigg], \notag\\
&C_7 = \dfrac{M}{2}\dfrac{\omega_1-\omega_2}{m_1\omega_2+m_2\omega_1}\bigg[f_{c5}-\dfrac{\omega_1+\omega_2}{m_1+m_2}f_{c6}\bigg], \notag\\
&C_8 = \dfrac{M}{2}\dfrac{m_1+m_2}{m_1\omega_2+m_2\omega_1}\bigg[-f_{c5}+\dfrac{\omega_1+\omega_2}{m_1+m_2}f_{c6}\bigg]. 
\end{align}
$f_{ci}$ ($i=3,4,5,6$) are the radial wave functions of the $2^{++}$ state.

The positive energy part of the wave function of the $1^-$ ($^3S_1$) state has the form \cite{Wang:2005qx}:
\begin{align}
\varphi^{++}_{1^-}(q^\prime_\bot) &= (q^\prime_\bot \cdot \varepsilon_f) \quad \bigg[ D_1(q^\prime_\bot) + \dfrac{\slashed{P}_f}{M_f} D_2(q^\prime_\bot) + \dfrac{\slashed{q}^\prime_\bot}{M_f} D_3(q^\prime_\bot) + \dfrac{\slashed{P}_f \slashed{q}^\prime_\bot}{M_f^2} D_4(q^\prime_\bot) \bigg] \notag
\\
&\quad+ M_f \slashed{\varepsilon}_f \quad \bigg[  D_5(q^\prime_\bot) + \dfrac{\slashed{P}_f}{M_f} D_6(q^\prime_\bot) + \dfrac{\slashed{q}^\prime_\bot}{M_f} D_7(q^\prime_\bot) + \dfrac{\slashed{P}_f \slashed{q}^\prime_\bot}{M_f^2} D_8(q^\prime_\bot) \bigg] ,
\end{align}
where
\begin{align}
& D_1 = \dfrac{1}{2 M_f (m_1 \omega_2 + m_2 \omega_1)} \big[ (\omega_1 + \omega_2)q^{\prime2}_\bot f_{d3} + (m_1 + m_2)q^{\prime2}_\bot f_{d4} + 2 M_f^2 (\omega_2 f_{d5} - m_2 f_{d6} ) \big], \notag \\
& D_2 = \dfrac{1}{2 M_f (m_1 \omega_2 + m_2 \omega_1)} \big[ (m_1 - m_2) q^{\prime2}_\bot f_{d3} + (\omega_1 - \omega_2q^{\prime2}_\bot f_{d4} - 2 M_f^2 (m_2 f_{d5} - \omega_2 f_{d6} ) \big], \notag \\
& D_3 = \dfrac{1}{2} \bigg[ f_{d3} + \dfrac{m_1 + m_2}{\omega_1 + \omega_2} f_{d4} - \dfrac{2 M_f^2}{m_1 \omega_2 + m_2 \omega_1} f_{d6} \bigg], \notag \\
& D_4 = \dfrac{1}{2} \bigg[\dfrac{\omega_1 + \omega_2}{m_1 + m_2} f_{d3} + f_{d4} - \dfrac{2 M_f^2}{m_1 \omega_2 + m_2 \omega_1} f_{d5} \bigg], \notag \\
& D_5 = \dfrac{1}{2} \bigg[ f_{d5} - \dfrac{\omega_1 + \omega_2}{m_1 + m_2} f_{d6} \bigg] , \qquad\qquad D_6 = \dfrac{1}{2} \bigg[- \dfrac{m_1 + m_2}{\omega_1 + \omega_2} f_{d5} + f_{d6} \bigg], \notag \\
& D_7 = \dfrac{M_f}{2} \dfrac{\omega_1 - \omega_2}{m_1 \omega_2 + m_2 \omega_1} \bigg[ f_{d5} - \dfrac{\omega_1 + \omega_2}{m_1 + m_2}  f_{d6}\bigg], \notag \\
& D_8 = \dfrac{M_f}{2} \dfrac{m_1 + m_2}{m_1 \omega_2 + m_2 \omega_1} \bigg[ - f_{d5} + \dfrac{\omega_1 + \omega_2}{m_1 + m_2}  f_{d6}\bigg].
\end{align}
$f_{di}$ ($i=3,4,5,6$) denotes the radial wave functions of the $1^{-}$ state.

In our expression above, the definition $\omega_i = \sqrt{m_i^2-q_{i\bot}^2}$ ($\omega_i = \sqrt{m_i^2-q^{\prime2}_{i\bot}}$ in the wave functions of the final meson $J/\psi$) is used. 

\bibliography{references.bib} 

\end{document}